\documentclass[aps,prl,preprint,groupedaddress]{revtex4}
\usepackage{graphicx}

\begin{document}

\title{Statistical Mechanics of finite arrays of coupled bistable elements}

\author{Jos\'e G\'omez-Ord\'o\~nez, Jos\'e M. Casado and  Manuel Morillo}
\email{morillo@us.es}
\affiliation{Universidad de Sevilla. Facultad de F\'{\i}sica. \'Area de
F\'{\i}sica Te\'orica. Apartado de Correos 1065. Sevilla 41080. Spain}
\author{ Christoph Honisch and Rudolf Friedrich}
\affiliation{Institut f\"ur Theoretische Physik. Westf\"alische Wilhelms-Universit\"at M\"unster, D-48149 M\"unster, Germany}

\date{\today}

\begin{abstract}
We discuss the equilibrium of a single collective variable characterizing a finite set of coupled, noisy, bistable systems as the noise strength, the size and the coupling parameter are varied. We identify distinct regions in parameter space. The results obtained in prior works in the asymptotic infinite size limit are significantly different from the finite size results. A procedure to construct approximate 1-dimensional Langevin equation is adopted. This equation provides a useful tool to understand the collective behavior even in the presence of an external driving force.    
\end{abstract}


\maketitle

Stochastic resonance (SR) is a phenomenon where the response of a nonlinear dynamical system to an external driving is enhanced by the action of noise \cite{rmp}. Although SR has been mainly discussed for simple systems, its analysis has also been extended to complex systems with many interacting units \cite{lindner,schi,neiman}. Our present study is prompted by recent studies on  SR in finite arrays \cite{Pikovsky,us06,us08,cubero}. Typically,  
the noise strength is the parameter varied to observe SR effects \cite{rmp}. In arrays, the coupling strength \cite{schi,us06}, as well as the system size \cite{neiman, Pikovsky} have also been considered as  parameters leading to SR. The complexity of the SR analysis in arrays can be facilitated by an adequate understanding of the equilibrium state. In this work, we carry out a detailed numerical analysis of the equilibrium distribution of the collective variable. The reduction of the multidimensional Langevin dynamics to an effective 1-dimensional Langevin equation greatly simplifies the analysis of the system response to weak driving forces \cite{Pikovsky,cubero}. Here we will assess the possibility of such a reduced Langevin description. 

Our model consists of a set of $N$ identical bistable units, each of them
characterized by a variable $x_i(t)\, (i=1,\ldots,N)$ satisfying a
stochastic evolution equation (in dimensionless form) of the type
\cite{deszwa,dawson,Pikovsky}
\begin{equation}
\dot{x}_i=x_i-x_i^3+\frac{\theta}{N}\sum_{j=1}^N(x_j-x_i)+\sqrt{2D}\xi_i(t),
\label{eq:lang}
\end{equation}
where $\theta$ is a coupling parameter and the term $\xi_i(t)$ represents a white noise with zero average and
$\left\langle \xi_i(t) \xi_j(s) \right\rangle = \delta_{ij}\delta
(t-s)$. 
The set $x_i(t)$ is an $N$-dimensional Markovian process.

We are interested in the properties of a collective variable, $S(t)=\frac 1N \sum_j x_j(t)$,
characterizing the chain as a whole.  Even though the set $x_i(t)$ is an N-dimensional
Markovian process, $S(t)$ in general is not.  Consequently, there is no reason why $S(t)$ should satisfy a Langevin equation. The equilibrium probability density of the collective variable is
$P^\mathrm{eq}(s)= \langle \delta(s-S(t)) \rangle^\mathrm{eq}$,  
where the average is taken with the $N$-dimensional equilibrium distribution for the $x_i(t)$ process. 
An exact analytical evaluation of the multidimensional integral is, in general, impossible. 

\begin{figure}
\includegraphics[width=8cm]{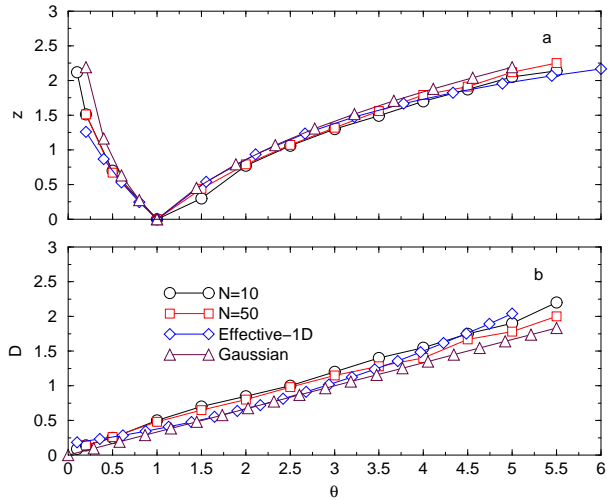}%
\caption{\label{uno} (Color online)Numerically determined lines separating different regions in parameter space. In the $z$ \textit{vs.} $\theta$ plane (a) $P^\mathrm{eq}(s)$ is bimodal above the line and monomodal below it. In the $z$ \textit{vs.} $D$ plane (b), $P^\mathrm{eq}(s)$ is monomodal above the line or multimodal below it.
 The circles correspond to a system with $N=10$, while the squares are for $N=50$. The coefficient $b$ of the effective Langevin equation, Eq.\ (\ref{effec1}) does not exist above the line with diamonds. The line with triangles correspond to the Gaussian approximation $D=3\theta$ in \cite{deszwa} for an infinite size system. }
\end{figure}

In the limit $ N \rightarrow \infty$, Desai and Zwanzig \cite{deszwa}
carried out an asymptotic analysis of the equilibrium density
$P^\mathrm{eq}_\infty(s)$ based on a saddle point approximation. Their analysis shows that
 $P^\mathrm{eq}_\infty(s)$ remains non-Gaussian 
even if the limit $\theta \rightarrow 0$ is taken at the end. A convenient
parameter is $z=\frac {|\theta-1|}{\sqrt{2D}}$. The asymptotic $P^\mathrm{eq}_\infty(s)$
 is not necessarily unique. Indeed, there is a
critical line in a $z$ \textit{vs.} $\theta$ diagram such that for $z$ values
below the critical line, there is a single stable $P^\mathrm{eq}_\infty(s)$
which is symmetrical around $s=0$ and either bimodal (for $\theta <
1$) or monomodal (for $\theta > 1$). On the other hand, for any $z$
value above the critical line, there are two stable coexisting monomodal
distributions centered around values $\pm s_0$, plus one unstable distribution centered at $s=0$.  Desai
and Zwanzig also discuss a Gaussian approximation to the general saddle point expression leading to a critical line $\theta=3D$. Points on this critical line in the $z$
\textit{vs.} $\theta$ plane are
depicted  in Fig.\ (\ref{uno}a) and in the $D$ \textit{vs.} $\theta$ plane in Fig.\ (\ref{uno}b) (triangles).

For finite systems, we rely on numerical simulations to obtain
information on $P^\mathrm{eq}(s)$. The set of equations in
Eq.\ (\ref{eq:lang}) are numerically integrated for very many
realizations of the noise terms. After a transient period we start gathering
data and average over the noise realizations to construct histograms estimating $P^\mathrm{eq}(s)$.  We
also evaluate the equilibrium time correlation function of the
collective variable and check that the system has indeed relaxed to an
equilibrium situation. The H-theorem for finite systems guarantees \cite{risken} that the equilibrium distribution
function of the $N$-dimensional $x_i(t)$ process is unique regardless of the initial condition.
Thus, it follows that $P^\mathrm{eq}(s)$ also exists and it
is unique. Nonetheless, its functional form might depend on the
parameter values considered.

\begin{figure}
\includegraphics[width=8cm]{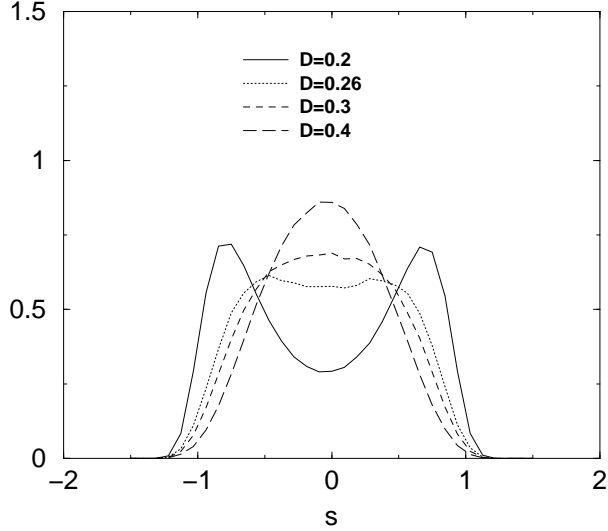}%
\caption{\label{dos} $P^\mathrm{eq}(s)$ for a $N=10$ units system with $\theta=0.5$ and several values of $D$, obtained from the simulations of the full set of equations, Eq.\ (\ref{eq:lang})}
\end{figure}

\begin{figure}
\includegraphics[width=7cm]{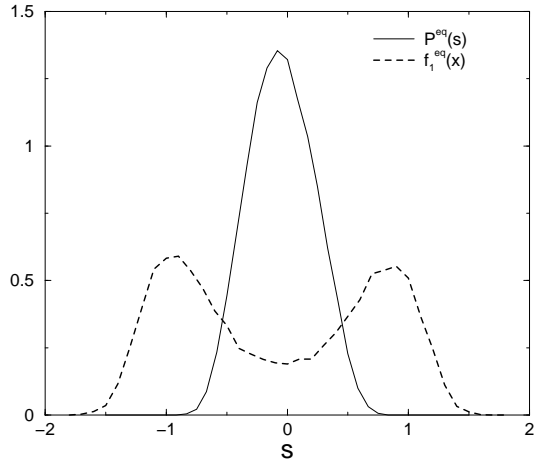}%
\caption{\label{tres} Plots of the equilibrium distribution for the
collective variable $P^\mathrm{eq}(s)$ and the equilibrium distribution
for one of the subsystems, $f^\mathrm{eq}_1(x)$, for a system with $N=50$
units coupled with a coupling strength $\theta=0.2$ and with $D=0.15$.}
\end{figure}

Our numerical findings indicate that there exists a line separating
different regions in parameter space. In  Fig.\ (\ref{uno}a) we depict the line for $N=10$ (circles), and  for $N=50$ (squares) in a $z$ \textit{vs.} $\theta$
plot. For points above the line, $P^\mathrm{eq}(s)$  has a local minimum at $s=0$ separating two maxima
symmetrically located around zero. Below the line, $P^\mathrm{eq}(s)$ has a single maximum at $s=0$.
In the $D$ \textit{vs.} $\theta$ plot in Fig.\ (\ref{uno}b), the
equilibrium probability density for parameter values above the
depicted line is always monomodal, with a maximum at $s=0$, while it is multimodal for points below the line. An example of this exchange of shape as $D$ is varied with $\theta$ kept constant 
can be seen in Fig.\ (\ref{dos}) for 
$\theta=0.5$ and $N=10$. The
transitions among barriers in the $N$-dimensional energy surface are
induced by the noise. Then, for large noise strengths, the
random trajectories of $x_i(t)$ have good chances of exploring all the
attractors quite frequently with numerous jumps over the barriers,
leading to a single maximum distribution for the global variable. For
low noise strengths the bimodality of the distribution is to be expected with more scarce jumps over high energy regions. It should be pointed out that the shape of $P^\mathrm{eq}(s)$ does not imply the shape for the one-particle equilibrium distribution function $f^\mathrm{eq}_1(x)$ obtained by integrating the joint
 equilibrium probability distribution over all the degrees of
 freedom except one. An example is shown in Fig.\ (\ref{tres}), where we note
 that for the parameter values $N=50$, $D=0.15$, and $\theta=0.2$,
 $P^\mathrm{eq}(s)$ is monomodal, while $f^\mathrm{eq}_1(x)$ is
 bimodal.

The features just described for finite systems are in sharp contrast
with the behavior found in previous works within the $N \rightarrow
\infty$ limit \cite{dawson}. The existence of two possible equilibrium
distributions above the critical line is impossible in finite
systems. Also, in the asymptotic infinite $N$ limit, low values of $z$
favor a single bimodal distribution for $\theta < 1$, or monomodal for $\theta > 1$,
while for finite values of $N$ the distribution $P^\mathrm{eq}(s)$ is
monomodal for all $\theta$. On the other
hand, the two possible single maximum stable equilibrium distributions
found for large values of $z$ in the asymptotic limit are replaced by
a single bimodal distribution in the finite size case.

It should also be pointed out that the lines separating the different
regions in either the $z$ or $D$ \textit{vs.} $\theta$ plots are quite
insensitive to the $N$ value used, as long as it remains finite. Thus,
the change in the shape of the global equilibrium
distribution does not seem to depend much on the system size, as long
as it remains finite.

In the asymptotic limit ($ N \rightarrow \infty$), Desai and Zwanzig
also showed that the dynamics could be casted in terms of a truly
nonlinear Fokker-Planck equation (i. e., nonlinear in the probability
distribution) consistent with the bifurcation of the equilibrium
probability distribution. Further details about the description of the
system in the $N \rightarrow \infty$ limit were discussed later by
analytical studies or numerical simulations
\cite{dawson}. There have been several
attempts to construct effective $1$-dimensional Langevin dynamics for
the collective variable for finite systems\cite{Pikovsky,cubero}. Although such
an equation does not necessarily exist as $S(t)$ is not necessarily a
Markovian process, when it does, it might be a useful and reliable approximation. In
\cite{Pikovsky}, Pikovsky et al. used the Gaussian truncation of an infinite
hierarchy of equations for the cumulants and a slaving principle to
construct an effective $1$-dimensional Langevin equation. Its explicit form is
\begin{equation}
\label{effec1}
 \dot S = aS-bS^3+\sqrt{\frac {2D}N}\chi(t),
\end{equation}
with $\left\langle \chi(t) \chi(s) \right\rangle = \delta
(t-s)$ and the coefficients $a$ and $b$ given by
\begin{eqnarray}
 a&=&1+0.5(\theta-1)-0.5\sqrt{(\theta-1)^2 + 12 D}; \nonumber \\
 b&=&\frac {4a}{2-\theta + \sqrt{(2+\theta)^2-24D}}.
\end{eqnarray}
Unfortunately,
the drift term appearing in such equation is a complex number for $D$
values below the line labelled as ``effective 1-D'' in the $z$-$\theta$ plane in Fig.\ (\ref{uno}a) (or above the corresponding line in the $D$-$\theta$ plane in Fig.\ (\ref{uno}b))
and so the effective Langevin equation, Eq.\ (\ref{effec1}) does not exist for all values in parameter space.

The method described in \cite{siegert98pla} opens up 
another possibility to construct an effective 1-dimensional Langevin equation by numerically estimating the
Kramers-Moyal coefficients $D^{(n)}(x,t)$ for a given data set of a general Markovian process $X(t)$. These coefficients are defined as
\begin{equation}
\label{coeff}
D^{(n)}(x,t) =\frac{1}{n!} \lim_{\tau \rightarrow 0} \frac{1}{\tau}
\left<(X(t+\tau) - x\right)^n>|_{X(t) = x}~.
\end{equation}
As described in \cite{siegert98pla} these conditional moments can be estimated for the smallest available $\tau$ by computing histograms. Given that the collective variable $S(t)$ is a Markovian process on this scale, i.e., the Markov length is smaller than $\tau$, this procedure yields a reliable 1-dimensional Langevin description of the form
\begin{equation}
\label{efec}
 \dot S = D^{(1)} (s,t) + \sqrt{2D^{(2)}(s,t)}\chi(t).
\end{equation}
We find that the fourth coefficient $D^{(4)}$ is about four orders of
magnitude smaller than $D^{(2)^2}$, i. e., it practically vanishes. Hence, the
Pawula theorem guarantees that the third and all higher coefficients do also
vanish.

\begin{figure}
 \resizebox{.5\textwidth}{!}{\rotatebox{270}{\includegraphics{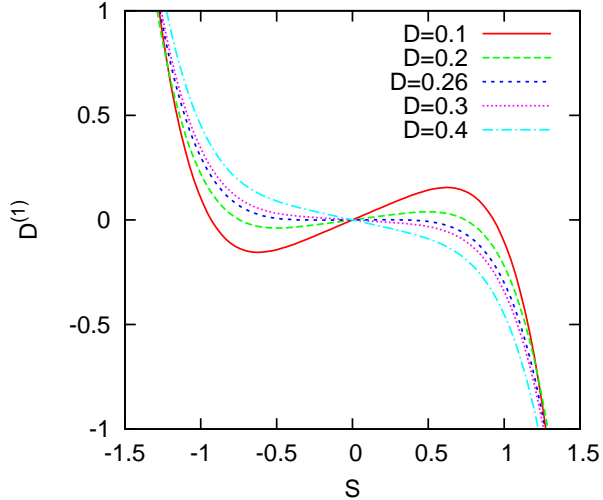}}}
 \caption{(Color online) The drift coefficient $D^{(1)}(s)$ for the effective Langevin, Eq.\ (\ref{efec}), for a system described by Eq.\ (\ref{eq:lang}) with  $N=10$, $\theta=0.5$ and several $D$ values }
 \label{dr2}
\end{figure}

In Fig. \ref{dr2} we depict our numerically estimated $D^{(1)}(s)$ for the system in Eq.\ (\ref{eq:lang}) with  $N=10$, $\theta=0.5$ and varying noise strengths. The noise value $D_c\approx0.26$ corresponds to the transition point in the $D$ \textit{vs.} $\theta$ diagram. The drift coefficients can be fitted with 
fifth degree odd polynomials, in contrast to the result derived by Pikovsky et al.
The diffusion coefficient is practically constant and satisfies
\begin{equation}
\label{difu}
 D^{(2)} = \frac{D}{N}\,.
\end{equation}
This relation is also valid for all other parameters we have tested.

\begin{figure}
\includegraphics[width=8cm]{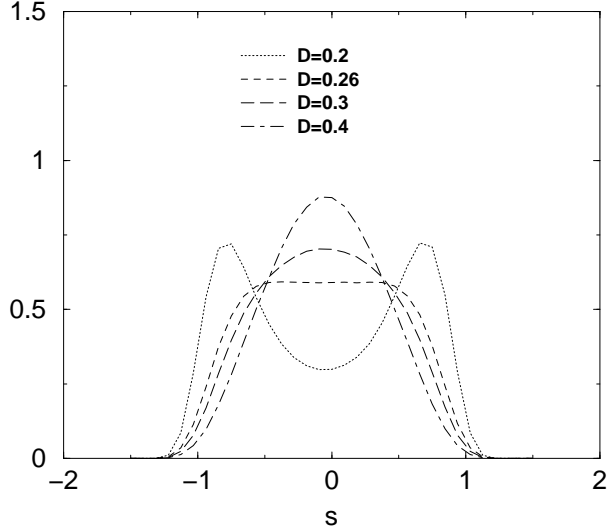}%
\caption{\label{peqeffec} Equilibrium distribution functions for the collective variable of a $N=10$ units system with coupling parameter $\theta=0.5$ and several $D$ values obtained from the approximate Langevin equation  Eq.\ (\ref{efec}).}
\end{figure}

We have also generated histograms for the equilibrium distribution using the effective Langevin equation Eq.\ (\ref{efec}) for parameter values above and below the transition line in Fig.\ (\ref{uno} b). The results are shown in Fig.\ (\ref{peqeffec}). Comparing with Fig.\ (\ref{dos}) we see that the  effective Langevin equation in Eq.\ (\ref{efec}) reproduces quite correctly the results obtained from the full set of equations for points above and below the transition line.   

\begin{figure}
\includegraphics[width=7cm]{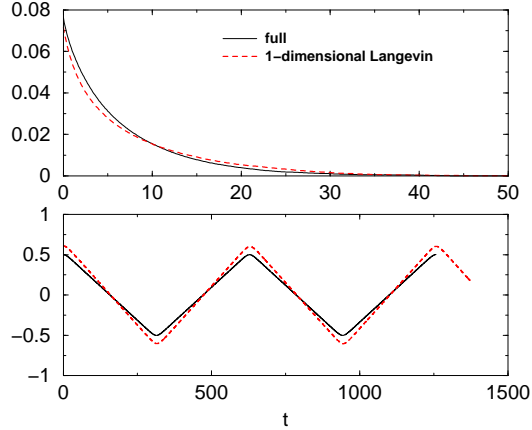}%
\caption{\label{corre} (Color online) The coherent (lower panel) and incoherent (upper panel) parts of the correlation function of $S(t)$ in a system with $N=10$, $\theta=0.5$, $D=0.24$ driven by a dichotomic force with $A=0.1$ and $\Omega=0.01$. The results depicted in black correspond to the simulations of Eq.\ (\ref{eq:lang}), while the red dashed lines are the results obtained with the 1-dimensional Langevin equation in (\ref{efec}).}
\end{figure}

In previous work \cite{us08}, we have analyzed the phenomenon of SR for the collective variable of an array driven by weak periodic forces. It was demonstrated that when the noise strength is varied while keeping $\theta$ and $N$ constant, the signal-to-noise ratio (SNR) of $S(t)$ 
shows a nonmonotonic behavior with $D$. Even for weak driving amplitudes, the SNR reaches values much larger than those typically observed in single unit systems for the same driving forces. This enhancement is largely due to the strong reduction of the fluctuations in the driven system with respect to those present in the absence of driving. The multidimensional character of the full potential relief makes it difficult to give a simple explanation of the SNR enhancement. The simplicity of the 1-dimensional Langevin equation allows us to understand the reduction of the fluctuations in terms of the drastic differences between the drift coefficient $ D^{(1)}(s)$ and $D^{(1)} (s,t)$ in driven systems.  For driven systems, Eq.\ (\ref{efec}) also leads to a good approximation to the correlation function as seen in Fig.\ (\ref{corre}) where we depict the results for the incoherent part (upper panel) and the coherent part (lower panel) of the correlation function for $S(t)$ as obtained from the simulation of Eq.\ (\ref{eq:lang}) (solid lines) and the approximate Langevin equation, Eqs.\ (\ref{efec}) and (\ref{coeff}) (dashed lines) for $N=10$, $\theta=0.5$, $D=0.24$ and a driving dichotomic force with amplitude $A=0.1$ and fundamental frequency $\Omega=0.01$. $D^{(2)} $ is well approximated by (\ref{difu}) while $D^{(1)}(s,t) $ is fitted by fifth order polynomials for each half-period. 

In conclusion, we have analyzed  a single collective variable characterizing a finite set of noisy bistable units with global coupling. We find several regions in  parameter space separated by transition lines. The shape of $P^\mathrm{eq}(s)$ switches from monomodal to multimodal as we move across the transition line. There are relevant qualitative differences with the results obtained in the infinite size limit, even though the lines separating different regions in parameter space look quite similar. We have also found approximate 1-dimensional Langevin equations for $S(t)$. The change in the shape of $P^\mathrm{eq}(s)$  implies the change on the drift coefficient as the parameter values are modified. In the presence of driving, the corresponding 1-dimensional Langevin equation provides a reliable tool to understand the enhancement of the stochastic resonance effect in arrays relative to the one observed in a single bistable unit.

\begin{acknowledgments}
We acknowledge the support of the Ministerio de Ciencia e Innovaci\'on
of Spain (FIS2008-04120)
\end{acknowledgments}

\end{document}